\documentclass[useAMS,usenatbib]{mn2e}
\usepackage{graphicx}
\usepackage{color}
\usepackage{amssymb}

\def\degr{\hbox{$^\circ$}}
\def\arcmin{\hbox{$^\prime$}}

\def\fs{\hbox{$.\!\!^{\rm s}$}}
\def\fdg{\hbox{$.\!\!^\circ$}}
\def\farcm{\hbox{$.\mkern-4mu^\prime$}}
\def\farcs{\hbox{$.\!\!^{\prime\prime}$}}

\title[Peculiar Radio Structures in Abell\,585.]{Peculiar Radio Structures in the Central Regions\\ of Galaxy Cluster Abell\,585}
\author[M. Jamrozy et al.]
{M. Jamrozy$^{1}$\thanks{E-mail: \texttt{jamrozy@oa.uj.edu.pl}}, {\L}. Stawarz$^{2,\,1}$, V. Marchenko$^{1}$, A. Ku\'zmicz$^{1}$, M. Ostrowski$^{1}$,
\newauthor
C.C. Cheung$^{3}$, and M. Sikora$^{4}$ 
\\
$^{1}$Astronomical Observatory, Jagiellonian University, ul. Orla 171 , 30-244 Krak\'ow, Poland\\
$^{2}$Institute of Space and Astronautical Science, JAXA, 3-1-1 Yoshinodai, Chuo-ku, Sagamihara, Kanagawa 252-5210, Japan\\
$^{3}$Space Science Division, Naval Research Laboratory, Washington, DC 20375-5352, USA\\
$^{4}$Nicolaus Copernicus Astronomical Center, ul. Bartycka 18, 00-716 Warszawa, Poland}

\begin{document}

\date{}

\pagerange{\pageref{firstpage}--\pageref{lastpage}} \pubyear{0000}

\maketitle

\label{firstpage}

\begin{abstract} 
In this paper we analyze the peculiar radio structure observed across the central region of the galaxy cluster 
Abell\,585 ($z=0.12$). In the low-resolution radio maps, this structure appears uniform and diffuse on angular scales of 
$\sim 3$\arcmin, and is seemingly related to the distant ($z=2.5$) radio quasar B3\,0727+409 rather than to the cluster itself. 
However, after a careful investigation of the unpublished archival radio data with better angular resolution, we resolve the 
structure into two distinct arcmin-scale features, which resemble typical lobes of cluster radio galaxies with no obvious connection
to the background quasar. We support this conclusion by examining the spectral and polarization properties of the features, 
demonstrating in addition that the analyzed structure can hardly be associated with any sort of a radio mini-halo or relics of the cluster. 
Yet at the same time we are not able to identify host galaxies of the radio lobes in the available optical 
and infrared surveys. We consider some speculative explanations for our findings, including gravitational wave recoil kicks of 
SMBHs responsible for the lobes' formation in the process of merging massive ellipticals within the central parts of a 
rich cluster environment, but we do not reach any robust conclusions regarding the origin of the detected radio features.
\end{abstract}

\begin{keywords}
radiation mechanisms: non-thermal --- galaxies: active --- galaxies: clusters: individual: Abell\,585 --- galaxies: jets --- 
quasars: individual: B3\,0727+409 --- radio continuum: galaxies
\end{keywords}

\section{Introduction}

During merging processes leading to the formation of clusters of galaxies, large amounts of gravitational energy are released 
on timescales of the order of $\sim$\,Gyr. Most of this energy is contained in hot (temperatures $kT \lesssim 10$\,keV)
X-ray--emitting plasma which constitutes, along with the dark matter, the dominant fraction of the intracluster medium (ICM; e.g., \citealt{sa1}).
In addition to the thermal gas, however, $\sim \mu$G magnetic fields and ultrarelativistic electrons are present within the ICM as well, manifesting 
most clearly in extended diffuse radio structures such as giant and mini radio halos in the central parts of clusters, or radio relics at cluster 
peripheries (see the reviews by \citealt{ct02,ferrari08,feretti1}). These non-thermal constituents of
the ICM are believe to be related to the energy dissipation processes enabled by large-scale shocks formed at the outskirts of
merging systems, and/or turbulence driven by various mechanisms at early post-merger stages of a cluster lifetime. 
The presence of (or rather the amount of) hadronic cosmic rays in the diffuse cluster environment is still an open question, being currently
probed with new-generation $\gamma$-ray instruments (see, e.g., \citealt{ah09,ac10,al10}).

Besides radio halos and relics, extended lobes and plumes of radio galaxies make up yet another class of diffuse non-thermal 
structures often found in the ICM. These are formed due to the jet activity of accreting supermassive black holes (SMBHs) hosted by
the cluster galaxies. The large-scale morphologies of such lobes are rarely regular or symmetric, unlike in the case of `classical doubles'
typical for poorer (galaxy group) environments. Instead, lobes of cluster radio galaxies are often bent (`tailed' radio sources), irregular, 
or even amorphous, reflecting the dramatic impact of high-pressure ambient medium, high peculiar velocities of parent galaxies, 
and highly intermittent jet activity of cluster SMBHs (e.g., \citealt{mil80}; see also the discussion in \S\,5.3 below). Such structures may 
therefore be considered as useful probes of the dynamical state and structure of the ICM, or even as effective tracers of galaxy clusters at 
moderate and high redshifts (see, e.g., \citealt{mao1}). Since they are best characterized at low radio frequencies, the operation of the 
next generation of low-frequency interferometers like the LOw Frequency ARray \citep{vH13} and the Murchison Widefield Array \citep{tin13} are in this context much anticipated.

Recently, much attention has been given to radio sources located within the central parts of galaxy clusters, since the mechanical 
energy output of radio-loud active galactic nuclei (AGN) which are hosted by central elliptical (cD) galaxies is widely believed to be responsible for 
suppressing the cluster cooling flows, quenching the star-formation in cD systems and the growth of their SMBHs (see, e.g., \citealt{fa94,be04,mcn07}). In order to investigate the details of such feedback processes, a precise spectral and 
morphological characterization of central radio structures and of their surroundings is however obligatory, and this requires 
deep high-resolution exposures with various instruments operating from radio to X-ray bands. Without such extensive dataset,
the physics of the lobes-ICM interactions may remain elusive.

In this paper we analyze the peculiar radio structure observed in the direction of the central parts of the galaxy cluster Abell\,585 ($z=0.12$).
The paper is organized as follows. The galaxy cluster Abell\,585 is briefly introduced in \S\,2. Multi-frequency data for the system 
are analyzed in \S\,3, and discussed further in \S\,4. The interpretation of the obtained observational results is presented in \S\,5, 
and the summary of the studies is given in \S\,6. In the paper we assume modern cosmology with $H_0=71$\,km\,s$^{-1}$\,Mpc$^{-1}$, 
$\Omega_{\rm M}=0.27$, and $\Omega_{\Lambda}=0.73$ \citep{s1}. All source positions are given in the J2000.0 coordinate system.

\section{Galaxy Cluster Abell\,585}

The cluster of galaxies Abell\,585 (another designation ZwCl\,0727.1+4057; \citealt{z1}) is centered at 
$\alpha$\,$= \rm 07^{h}30^{m}57^{s}$ and $\delta$\,$= +40\degr 50\farcm7$ (the corresponding Galactic longitude and latitude of 
177\fdg88 and +24\fdg44, respectively; \citealt{a2}). The cluster is classified as a (B--M) type II system. The red magnitude 
of the cluster's tenth brightest member galaxy is 17.0\,mag, and there are in total 35 galaxies with apparent red magnitudes 
between mag$_{3}$ and mag$_{5}$. Abell\,585 occupies an area of 26\arcmin\ in diameter and its richness class is 0 \citep{a1}. 
Using the photometric redshifts of galaxies from the Sloan Digital Sky Survey Data 
Release 6 (SDSS DR6; \citealt{b1}), \citet{w1} identified a cluster named WHL\,J073045.4$+$405038, 
which overlaps with Abell\,585. The photometric redshift of this SDSS cluster is $z=0.121$, and the spectroscopic redshift of the 
brightest cluster galaxy ($r = 15.85$\,mag) is $z=0.1192$. The total $r$-band cluster luminosity summed over all 27 member galaxies, 
which are spread over $\sim 1$\,Mpc, is $\simeq 1.2 \times10^{12} L_{\odot}$.

In the catalogue of superclusters (SCL; \citealt{e1}), Abell\,585 together with Abell\,580 and Abell\,591 appear as a 
part of the SCL\,074, for which the central position is, approximately, [$\alpha$\,$= \rm 07^{h}36^{m}17^{s}$, $\delta$\,$= 41\degr59\farcm3$]. 
Abell\,580 is centered at [$\alpha$\,$= \rm 07^{h}25^{m}53^{s}$, $\delta$\,$=+41\degr25\farcm0$] at the redshift of $z=0.1118$, with the
diameter $\sim 30$\arcmin, (B--M) type II--III, and richness class 1. 
Abell\,591 is centered at [$\alpha$\,$= \rm 07^{h}42^{m}02^{s}$, $\delta$\,$=+43\degr57\farcm9$] at the redshift of $z=0.1178$, with the 
diameter of $\sim 34$\arcmin, morphology system of (B--M) type III, and richness class equal 1. 
The SCL\,074 triplet may therefore represent a rather dynamic merging structure.

\section{The Data}

The analysis presented in this paper is based on the archival optical, X-ray, and multi-frequency radio data, as described in detail below. 

\subsection{Optical Data}

\begin{figure*}
\includegraphics[width=0.85\linewidth, angle=0]{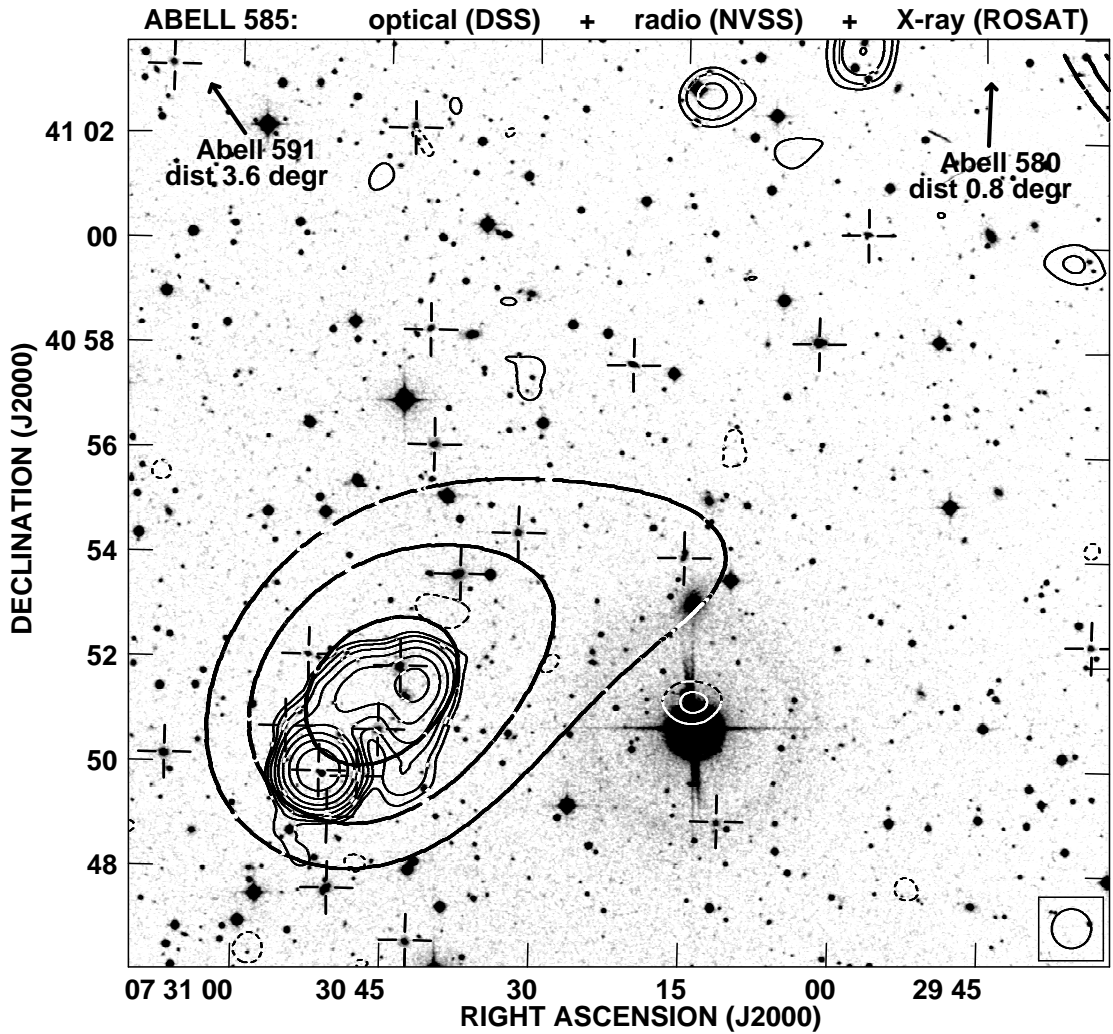}
\caption{Multi-band image of the central regions of the galaxy cluster Abell\,585. 
Optical $R$-band image from the DSS is shown in grey scale. The crosses mark the positions of the 
cluster member galaxies, as well as of the background quasar B3\,0727+409. 
The two arrows indicate the directions to the nearest clusters Abell\,580 and Abell\,591, 
along with the corresponding angular distances. The thin contours represent the 1400\,MHz radio emission 
measured in the NVSS survey. The contours are spaced by factors of $\sqrt2$ in brightness and plotted 
starting with 1.35\,mJy\,beam$^{-1}$. The dashed contours represent the negative flux-density. 
The size of the beam is shown by a circle in the bottom-right corner of the image. 
The thick contours denotes the $0.5-2.4$\,keV X-ray emission measured in the ROSAT survey. 
The data have been smoothed using a Gaussian kernel with $\sigma=3$\arcmin. The contours correspond to 
the level of photon counts\,pixel$^{-1}$ of 0.25, 0.29, and 0.55, respectively.}
\label{fig:sdss}
\end{figure*}

A Digital Sky Survey (DSS) image of the central part of the galaxy cluster Abell\,585 is shown in Figure\,\ref{fig:sdss}, 
where the cluster member galaxies, as well as the background quasar B3\,0727+409 (see Table\,1), 
are marked with the cross symbols. We use the Sloan Digital Sky Survey (SDSS; \citealt{gunn06}) database to locate galaxies and to establish their associations with the Abell\,585 cluster based on spectroscopic redshifts from the SDSS DR9 \citep{ahn1}. 
We defined all the member galaxies to have velocities in the range between $34,000$ and $38,000$ km/s (i.e. about $\pm$2000 km/s around the velocity corresponding to z=0.12). The redshifts of the member galaxies are provided in Table\,1 along with the source positions, $r$-band magnitudes, and central black hole masses estimated by us from the stellar velocity dispersion using the \citet{t1} relation except for the black hole mass for B3\,0727+409, estimated using the quasar emission lines (see \S\,5.1). In total, we found 21 galaxies within the 15\arcmin\, radius (corresponding to a projected linear size of about 4\,Mpc) around the central cluster region, plus a luminous 
face-on spiral with bright optical core lacking any spectroscopic redshift determination, but with the SDSS photometric redshift of $\sim 0.1$ and an optical apparent magnitude similar to the cluster member galaxies.

\begin{table*}
 \centering
 \begin{minipage}{130mm}
  \caption{Abell\,585 cluster member galaxies and the quasar B3\,0727+409.}
  \begin{tabular}{@{}ccclcc@{}}
  \hline
\hspace{0.5cm}$\alpha$(J2000.0)& $\delta$(J2000.0)           & $z$       &ID & $r$ &$\log(M_{\rm BH}/M_{\odot})$\\
\hspace{0.5cm}[h m s]          & [\degr\, \arcmin\, \arcsec] &           &   &[mag]&                            \\
\hspace{0.5cm}(1)              & (2)                         &(3)        &(4)& (5) &(6)                         \\
\hline

\hspace{0.5cm}07 30 51.35       &$+$40 49 50.8                & 2.500000$\pm$0.001000  & Q &18.645$\pm0.016$&8.52$\pm$0.29  \\  
                                &                             &                        &                    &               \\
\hspace{0.5cm}07 29 33.74       &$+$40 52 23.0                & 0.119155$\pm$0.000135  & G &17.534$\pm$0.015& 7.07$\pm$0.16 \\
\hspace{0.5cm}07 29 56.73       &$+$41 00 12.8                & 0.116985$\pm$0.000148  & G &17.368$\pm$0.035& 7.66$\pm$0.12 \\
\hspace{0.5cm}07 30 01.54       &$+$40 58 07.5                & 0.117351$\pm$0.000168  & G &15.995$\pm$0.019& 8.41$\pm$0.06 \\ 
\hspace{0.5cm}07 30 11.27       &$+$40 48 58.7                & 0.121375$\pm$0.000149  & G &17.853$\pm$0.016& 7.40$\pm$0.15 \\ 
\hspace{0.5cm}07 30 14.88       &$+$40 54 00.3                & 0.115038$\pm$0.000143  & G &17.905$\pm$0.029& 7.77$\pm$0.15 \\  
\hspace{0.5cm}07 30 20.25       &$+$40 57 40.3 	              & 0.116029$\pm$0.000150  & G &17.071$\pm$0.012& 7.92$\pm$0.10 \\
\hspace{0.5cm}07 30 31.61       &$+$40 54 26.4                & 0.117374$\pm$0.000172  & G &16.992$\pm$0.010& 8.08$\pm$0.09 \\  
\hspace{0.5cm}07 30 37.37       &$+$40 53 38.3                & 0.120159$\pm$0.000164  & G &16.652$\pm$0.050& 8.33$\pm$0.07 \\ 
\hspace{0.5cm}07 30 40.20       &$+$40 56 06.3                & 0.117546$\pm$0.000138  & G &17.128$\pm$0.017& 7.33$\pm$0.14 \\ 
\hspace{0.5cm}07 30 40.75       &$+$40 58 17.8                & 0.116965$\pm$0.000144  & G &17.404$\pm$0.011& 7.18$\pm$0.16 \\
\hspace{0.5cm}07 30 42.37       &$+$40 46 37.6                & 0.119720$\pm$0.000853  & G$^a$ &17.170$\pm$0.016& 7.21$\pm$0.16 \\ 
\hspace{0.5cm}07 30 42.59       &$+$41 02 09.1                & 0.117663$\pm$0.000161  & G &17.282$\pm$0.017& 8.10$\pm$0.11 \\ 
\hspace{0.5cm}07 30 42.62       &$+$40 51 17.7                & (0.1030$\pm$0.0149)$^{\star}$& G$^b$ &16.721$\pm$0.006&     \\ 
\hspace{0.5cm}07 30 43.28       &$+$40 51 52.1                & 0.118739$\pm$0.000166  & G$^c$ &16.521$\pm$0.026& 8.52$\pm$0.07 \\
\hspace{0.5cm}07 30 45.43       &$+$40 50 38.9                & 0.119675$\pm$0.000139  & G &17.009$\pm$0.012& 7.83$\pm$0.09 \\ 
\hspace{0.5cm}07 30 47.51       &$+$40 49 44.8                & 0.118659$\pm$0.000164  & G &17.333$\pm$0.015& 6.97$\pm$0.23 \\ 
\hspace{0.5cm}07 30 50.37       &$+$40 47 36.6                & 0.119227$\pm$0.000188  & G &16.184$\pm$0.016& 8.63$\pm$0.05 \\ 
\hspace{0.5cm}07 30 50.46       &$+$40 48 57.9                & 0.118476$\pm$0.000143  & G &17.826$\pm$0.016& 7.73$\pm$0.15 \\ 
\hspace{0.5cm}07 30 52.53       &$+$40 52 04.3                & 0.113726$\pm$0.000156  & G &17.829$\pm$0.026& 7.11$\pm$0.13 \\ 
\hspace{0.5cm}07 30 54.74       &$+$40 50 41.6                & 0.118776$\pm$0.000139  & G &17.475$\pm$0.016& 7.76$\pm$0.11 \\ 
\hspace{0.5cm}07 31 06.91       &$+$40 50 09.3                & 0.119612$\pm$0.000142  & G &17.533$\pm$0.012& 8.13$\pm$0.07 \\ 
\hspace{0.5cm}07 31 07.06       &$+$41 03 18.2                & 0.117050$\pm$0.000140  & G &17.481$\pm$0.017& 7.32$\pm$0.12 \\ 
\hline
\end{tabular}\\
(1-2) source position; (3) spectroscopic redshifts from SDSS DR9; (4) source identification (galaxy `G' or quasar `Q'); 
(5) $r$-band SDSS Petrosian magnitude (radius $= 1\farcs63$); (6) black hole mass.\\
($^{\star}$) only the photometric redshift determination is available;\\
($^a$) this galaxy hosts a radio-quiet AGN;\\ 
($^b$) bright face-on spiral marked with `S' in Figure\,\ref{fig:perley};\\
($^c$) bright elliptical marked with `P' in Figure\,\ref{fig:perley}.
\end{minipage}
\end{table*}

\subsection{X-ray Data}

In order to characterize the X-ray properties of the galaxy cluster Abell\,585, we analyze the data from 
the ROSAT All-Sky Survey (RASS; \citealt{v1,v2}), and in particular the data with seq.id RS931017N00 
from the PSPC instrument. We note that currently it is not possible to handle the complete (temporal and spectral) 
analysis of the scanning mode data from ROSAT, because each individual RASS field scanning-mode observation 
is a combination of some number of individual PSPC fields. Therefore we re-bin the data from the event file into 
two energy bands $0.11-0.52$\,keV and $0.52-2.47$\,keV, covering in this way the entire ROSAT range. 
We also chose the spatial binning with the pixel size of $\simeq 1$\arcmin\ (the binning factor of 120)
in order to enhance the signal-to-noise ratio. 

A weak elongated X-ray structure, with the position of the brightness peak overlapping with that of the highest surface 
brightness radio structure (see the next section), is pronounced in the high-energy ROSAT map 
(see Figure\,\ref{fig:sdss}). The $0.1-2.4$ keV flux of the entire ROSAT source is $(1.10\pm0.35)\times 10^{-12}$\,erg\,cm$^{-2}$\,s$^{-1}$
when corrected for the Galactic absorption in the direction of the source \citep{bri07}. This level of the X-ray emission is, in principle,
consistent with what could be expected from the unresolved core of the radio-loud quasar B3\,0727+409, taking into account its radio and 
optical fluxes. However, as noted above, the detected X-ray structure appears elongated, with the brightness peak shifted by 
about 2\arcmin\ north-west from the quasar position. We confirmed the reality of this offset by analyzing the encircled energy 
function and the PSF blurring effects. We emphasize in this context, however, that even though the astrometric 
accuracy of source positions for RASS data is about 7\arcsec\ \citep{p1}, the position of the X-ray maximum 
for the feature in question is estimated using the smoothed image which contains only a few source counts, and as 
such is rather uncertain.

In addition, we check the two observations made with the \emph{Swift}-XRT which overlapped with the 
galaxy cluster Abell\,585. The XRT data (with the total exposure of 2.2\,ksec; Target ID 35951) do reveal some enhanced X-ray emission 
in the region of interest, but extremely poor photon statistics precludes us from any detailed analysis.

\subsection{Radio Data}

During the last three decades the vicinity of Abell\,585 was mapped in five northern-sky radio interferometric surveys:
the National Radio Astronomy Observatory (NRAO) Very Large Array (VLA) Sky Survey at 1400\,MHz
(NVSS; \citealt{co1}), the Faint Images of the Radio Sky at Twenty-cm survey at 1400\,MHz
(FIRST; \citealt{b3}), the Westerbork Northern Sky Survey at 327\,MHz
(WENSS; \citealt{r1}), the 6th Cambridge Survey at 151\,MHz (6C; \citealt{h11}), 
and the VLA Low-Frequency Sky Survey at 74\,MHz (VLSS; \citealt{c2}). 
This area was also covered in the single-dish Green Bank 4850\,MHz northern-sky survey 
(GB6; \citealt{g1}). The NVSS map of the cluster reveals the bright point-like core of the 
background quasar B3\,0727+409 along with the arcmin-scale tail extending to the north-west 
from the quasar position in the direction of the cluster center (Figure\,\ref{fig:sdss}). 
The diffuse radio tail looks rather striking, and at first glance seems to represent the one-sided jet/lobe
of the background quasar rather than any cluster-related structure. However, those low-angular resolution radio 
maps are not sufficient to prove or disprove this apparent association. We therefore gathered and analyze 
all the other available radio data for the source, some with angular resolution superior to that of the NVSS. 
The observational details for radio data we used in the analysis are provided in Table\,2. 

\begin{table*}
\scriptsize
\caption{Radio data analyzed in this paper}
\begin{tabular}{r c l r c r ccc c}
\hline  
Frequency      & Instrument  & Observation   &Time     & Beam                       & rms                 &\multicolumn{3}{c}{Flux density}                              &Refs.  \\ 
               &             & date          &on source&                            &                     &\multicolumn{2}{c}{The quasar}         & Total structure     &       \\
 $[$MHz$]$     &             &               &[min]    &[arcsec$^2$]                & [mJy beam$^{-1}$]   &[mJy beam$^{-1}$]   &   [mJy]            &  [mJy]              &       \\ 
(1)            &    (2)      &    (3)        &(4)      & (5)                        &   (6)               & (7)                & (8)                &   (9)               &  (10) \\ 
\hline
               &             &               &         &                            &                     &                    &                    &                     &       \\
  73.8     &VLA-B            & 2003 Sep 20   &75       & 75  $\times$  75           &67.2                 &                    &                    &1103.5$\pm$20.5      &  1    \\
 151.5     &Cambridge        & 1977          &         & 252.0  $\times$  385.3     &40.0                 &                    &                    &1090.0$\pm$75.0$^{a}$&  2    \\   
 326.8     &WSRT             & 1991 Feb 11   &36       & 54.0  $\times$  82.6       &3.3                  &                    &                    &720.3$\pm$38.5       &  3    \\
 408.0     &Northern Cross   & 1977 Feb-May  &         & 156.0  $\times$  289.0     &10.0                 &                    &                    &563.5$\pm$20.0$^{b}$ &  4    \\
1400.0     &VLA-D            & 1993 Dec 15   &0.5      & 45  $\times$  45           &0.45                 &362.6$\pm$18.1      & 392.3$\pm$19.6     &539.7$\pm$27.1       &  5    \\ 
1400.0     &VLA-B            & 1997 Feb 20   &3        & 5.4  $\times$  5.4         &0.15                 &344.8$\pm$17.2      & 352.4$\pm$17.6     &                     &  6    \\
1425.0     &VLA-C            & 1994 Oct 25   &1.3      & 16.4  $\times$  14.9       &0.06                 &342.6$\pm$17.1      & 342.7$\pm$17.1     &                     & 7 \\
1464.9     &VLA-C            & 1993 Jul 09   &11.5     & 15.3  $\times$  12.5       &0.1                  &346.1$\pm$17.3      & 376.9$\pm$18.9     &                     & 7 \\
1489.9     &VLA-A/B          & 1991 Oct 13   &2.7      & 4.5  $\times$  1.5         &0.04                 &382.3$\pm$19.1      & 401.7$\pm$20.1     &                     & 7 \\
2268.3     &VLBA             & 1996 Jun 07   &5.4      & 0.007  $\times$  0.003     &0.5                  &421.1$\pm$21.1      & 446.1$\pm$22.3     &                     &  8    \\
4710.1     &VLA-B            & 1994 Sep 06   &2.5      & 2.4  $\times$  1.4         &0.02                 &414.2$\pm$20.7      & 414.7$\pm$20.7     &                     & 7 \\
4850.0     &NRAO93M          & 1987 Oct 15   &         & 216.4  $\times$  194.0     &5.4                  &                    &                    &466.0$\pm$27.1       &  9    \\
4995.5     &VLBA             & 1998 Feb 08   &10.5     & 0.002  $\times$  0.002     &0.2                  &241.2$\pm$12.1      & 310.1$\pm$15.5     &                     & 10    \\
8338.3     &VLBA             & 1996 Jun 07   &5.4      & 0.002  $\times$  0.001     &0.5                  &147.5$\pm$7.4       & 311.4$\pm$15.6     &                     &  8    \\
8439.9     &VLA-C            & 1994 Oct 25   &1.5      & 2.5  $\times$  2.3         &0.04                 &322.6$\pm$16.1      & 322.8$\pm$16.1     &                     & 7 \\
14939.9    &VLA-A/B          & 1991 Oct 13   &4.3      & 0.4  $\times$  0.2         &0.07                 &316.7$\pm$15.8      & 326.9$\pm$16.3     &                     & 7 \\
30000.0    &Torun32M         & 2005 Apr--Aug &         & 1.2  $\times$  1.2         &                     & 93.0$\pm$5.0       &                    &                     &  11   \\
43340.0    &VLA-B/C          & 2000 Mar 05   &4.4      & 0.5  $\times$  0.4         &0.06                 & 95.0$\pm$4.8       & 98.3$\pm$5.1       &                     & 7 \\
\hline
\end{tabular}
\begin{flushleft}

Notes: 
$^a$ the peak flux only available; $^b$ the original B3 flux density given in the CKL scale \citep{kel1} is multiplied by a factor of 1.129 relative to the common Baars scale according to \citet{b55}.\\
References:
(1) VLSS: \citealt{c2};
(2) 6C: \citealt{h11};
(3) WENSS: \citealt{r1}; 
(4) B3: \citealt{f1};
(5) NVSS: \citealt{co1}; 
(6) FIRST: \citealt{b3}; 
(7) this paper; 
(8) \citealt{b2};
(9) GB2: \citealt{g1};
(10) \citealt{h1};
(11) OCRA-p: \citealt{l1}.\\
\end{flushleft}
\end{table*}

The area of interest has been observed a number of times with the VLA in the past, due to the fact that the quasar B3\,0727+409 is 
a VLA phase calibrator often observed in a snapshot mode (usually a few minutes of integration time; for details see col.\,4 in Table\,2). Hence, for the purpose of our investigation, apart from the 
publicly available survey FITS maps (i.e. VLSS, WENSS, FIRST, NVSS and GB2), we analyze available raw unpublished data from the VLA archive (these data are marked with number `7' in col. 10 of Table\,2), reducing them using standard procedures with the AIPS package. After preliminary CLEANing of the maps with 
the routine IMAGR, several iterations of self-calibration were performed to improve the maps' quality. Finally, 
the maps were corrected for primary beam attenuation using the PBCOR task. All the resulting flux densities are given in the \citet{b55} scale. The flux densities obtained directly from the FITS maps (either the maps made for the purpose of this investigation,
or the maps taken from the surveys) with the Astronomical Image Processing System (AIPS) task TVSTAT or JMFIT (in the case of a point source) are assumed to have the absolute flux calibration errors of 5\%; the flux densities taken from the literature are assumed
to have the errors as determined in the corresponding references. In the case of the flux density values which 
had not been measured directly from the FITS maps (e.g. those of the extended structure) but estimated by other means, 
we propagate the errors of the directly measured flux densities.
The resulting flux densities (along with the errors) of the quasar and of the total structure (i.e., quasar $+$ extended radio tail) 
are listed in Table\,2. The quasar flux densities are peak and integrated values, in order 
to clarify any possible contamination from the diffuse emission. The flux density errors of the quasar are calculated 
by adding the flux measurement errors and the JMFIT errors in quadrature. 

The multi-frequency radio data collected here can be divided into three categories: i) those which resolve the quasar core as an isolated source and give proper values of its flux-density (see Table\,2, col. 7 and 8), ii) those which give a proper overall image and flux-densities of the entire `quasar + extended halo' structure but have little ability to distinguish individual components as distinct sources (see Table\,2, col. 9), and iii) those which have adequate angular resolution to separate and image properly both the quasar core and the extended features but may underestimate the flux densities of the extended components (VLA-B 1400\,MHz and VLA-C 1425\,MHz observations). The former class of data (i) were used to determine the radio spectrum of the quasar and to examine whether the source is variable or not (for detail see \S\,4.1). The characterized shape of the quasar continuum allowed us next to subtract the contribution of the quasar core from the total  `quasar + extended halo' fluxes evaluated using the second (ii) class of data. In this way we obtained the total spectrum of the extended structure. The latter class of data (iii) allowed us finally to depict morphological details of the extended structure. Furthermore, the VLBA data were used to investigate the small-scale structure of the quasar core (for details see \S\,5.1).

\section{Results}

\subsection{Overall Radio Structure}

\begin{figure*}
\centering
\includegraphics[width=0.85\linewidth]{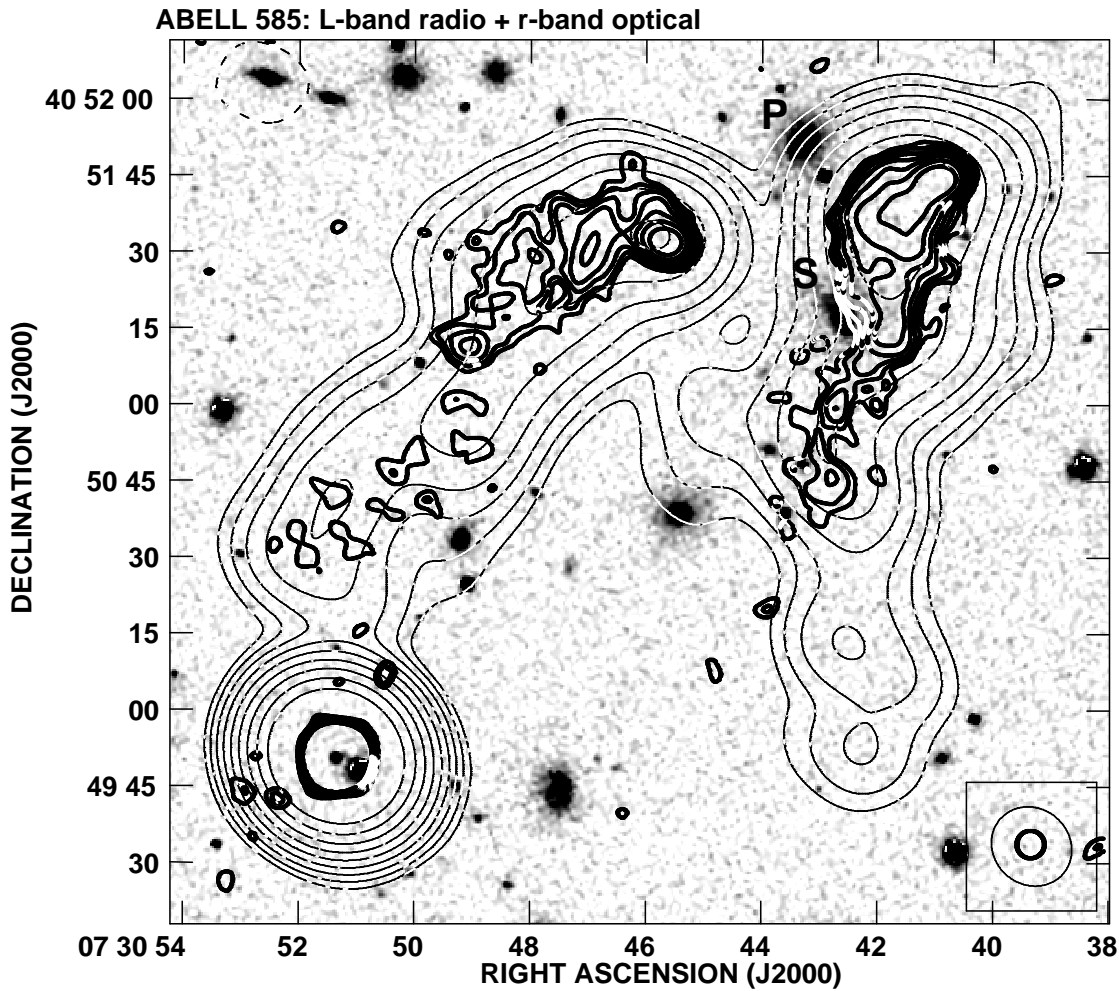}\\
\caption{L-band VLA images of the radio structures in Abell\,585 overlaid on 
the $r$-band optical field from the SDSS (DR9). {\bf Thin contours} correspond to the 1425\,MHz VLA C-array map of the entire source (see Table\,2 for details). The contour levels are spaced by a factor of $2$, and the first contour is 0.18\,mJy\,beam$^{-1}$. 
{\bf Thick contours} represent the VLA B-array map from the FIRST survey. The contour levels are spaced by a factor of $\sqrt2$, and 
starting from 0.35\,mJy\,beam$^{-1}$. The indices `S' and `P' denote the bright spiral galaxy and the particularly prominent elliptical
galaxy discussed in the text (see Table\,1). The size of the beams are given as ellipses in the bottom-right corner of the image.}
\label{fig:perley}
\end{figure*}

Figure\,\ref{fig:perley} presents our VLA image at 1425\,MHz of the entire radio source in Abell\,585 (the largest 
angular extent of about 2\farcm5) with an angular resolution of $\sim$15\arcsec\, along with the $\sim$5\arcsec\ 
angular resolution 1400\,MHz FIRST map. The diffuse feature seen in the NVSS map is now clearly resolved into 
two cometary-like structures with bright compact `heads' in their northern parts (hereafter the eastern one `E' and the western one `W'), and edge--dimmed tails elongated to the south. 
The high-frequency (e.g., 8439.9\,MHz) maps allow us to measure precisely the positions of the heads' flux maxima
as [$\alpha$\,$= \rm 07^{h}30^{m}45\fs78$, $\delta$\,$=+40\degr51\arcmin32\farcs4$] and [$\alpha$\,$=\rm 07^{h}30^{m}41\fs03$, 
$\delta$\,$=+40\degr51\arcmin44\farcs0$]. There are no optical counterparts at these positions in the DSS and SDSS maps. 

In order to measure the total flux density of the extended emission component (the two cometary-like structures), we first
measure the flux densities of the quasar core at different frequencies, and fit the spectrum with a single power-law function,
which returns $\log [S_{\nu}/{\rm mJy}] = - (0.077 \pm0.039) \times \log [\nu/{\rm MHz}] + (2.829 \pm 0.137)$.
Next we subtract the fitted quasar spectrum at a given frequency from the integrated flux of the entire structure, obtaining in this way
the spectrum of the extended structure only, $\log [S_{\nu}/{\rm mJy}] = - (0.408 \pm0.043) \times \log [\nu/{\rm MHz}] + 
(3.520 \pm 0.125)$. This estimate implies a total flux density of the extended structure at 1425\,MHz of about 171\,mJy, 
which is consistent with the total flux density measured in the VLA C--array map at 1425\,MHz for the western 
and eastern structures separately (see Figure\,\ref{fig:perley}, thick contours), i.e. $78.11\pm2.35$ and $48.29\pm1.46$\,mJy, 
respectively. A small disagreement here may be due to the fact that, as indicated by the multi-epoch radio data, the 
B3\,0727+409 core radio emission is variable, decreasing with time over the last decades, and this trend is particularly 
clear at 1.4, 5 and 8\,GHz. A spectral curvature for the extended component (see the next section) may also play a role.

\subsection{Radio Spectra}

The NVSS and WENSS images --- which have similar beam sizes and are sensitive to extended (up to a few arcmin) diffuse structures --- were used to create a spectral index map of Abell\,585. First, using the AIPS task SAD we subtracted the quasar B3\,0727+409 from both maps. 
Subsequently, the higher resolution map was convolved to the resolution of the 327\,MHz map. Since misalignment of the total-power map at two frequencies could produce systematic errors in the spectral-index map, we co-registered the positions of several bright point-like field sources which surround the target source on both maps. Further, both maps were brought to a common scale using the AIPS task HGEOM. Finally, the spectral index map was obtained using the AIPS task COMB. Regions with flux
density values below $3\times$rms were considered to be unreliable and blanked. The final grey-scale map of the spectral index is shown in Figure\,\ref{fig:pol}\,(a). The map indicates that there is a monotonic steepening of the radio spectrum of the diffuse component from north to south, i.e. from compact `heads' of the two resolved structures along their tails. The trend is not that clear in the case of the eastern tail because of an imperfect subtraction of a point source present around the tail's southern edge. Such a steepening is a classic spectral signature of an aging synchrotron-emitting plasma in the lobes and plums of radio galaxies, observed in almost all luminous classical doubles (for which the flattest radio spectra are seen at the positions of terminal hotspots, where the injection of fresh particles
takes place; e.g., \citealt{car91}) and also low-power radio galaxies including tailed sources in galaxy clusters (for which 
which the flattest radio spectra are seen around the jet base; e.g., \citealt{fer98}). 
The radio spectrum of the quasar core is inverted, as expected, with the 327-1400\,MHz spectral index of about $\alpha \simeq 0.31$ (defined here as $\alpha \propto \ln S_{\nu} / \ln \nu$ for the flux spectral density $S_{\nu}$). The mean spectral index over the extended 
structure is characteristic for lobes and plumes of radio galaxies in general, $\alpha \simeq -0.72$, and implies a relatively 
young population of the radio-emitting electrons ($<100$\,Myr). We note in this context that rather steep radio spectra seen at the south-western outskirts of the structure ($\alpha \sim -1.5$) may be due to a missing flux in the 1400\,MHz interferometric map only. Finally, using the VLA 4710.1 and 8439.9\,MHz maps we evaluate the spectral index of the eastern `head' as $-0.66 \pm 0.40$. 

\subsection{Polarization Properties}

\begin{figure}
\includegraphics[width=0.85\linewidth]{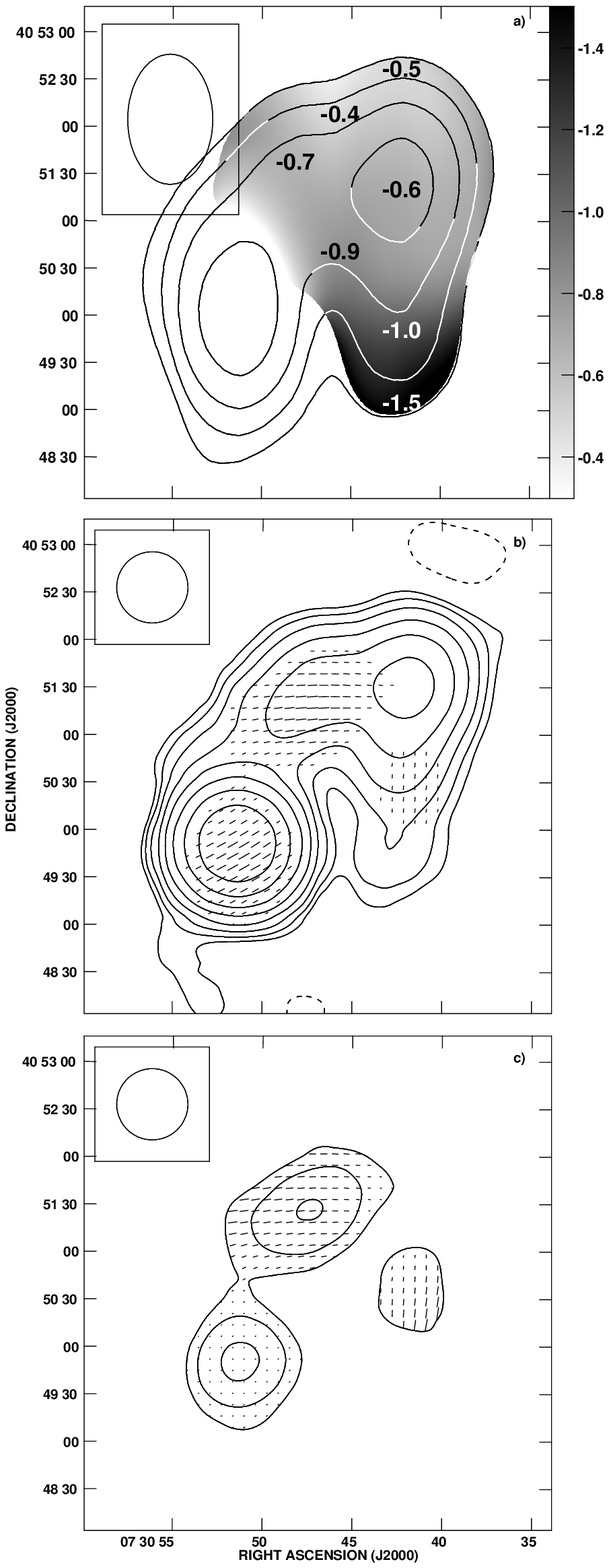}
\caption{{\bf a)} Spectral index map of the radio structure in Abell\,585 between 327\,MHz (WENSS) and 1400\,MHz (NVSS)
(grey scale) overlaid on the original WENSS map (contours spaced by a factor of 2 starting from
1.8\,mJy\,beam$^{-1}$). In several places of the figure spectral index values are given explicitly. 
 {\bf b)} Total intensity 1400\,MHz contours spaced by a factor of 2 
starting at 1.35\,mJy\,beam$^{-1}$, with the superimposed $E$-vectors, for which the 
length is proportional to the polarized intensity (10\arcsec\ corresponds to 5.5\,mJy\,beam$^{-1}$). 
{\bf c)} Linearly polarized 1400\,MHz intensity contours spaced by a factor of 2 starting from
0.9\,mJy\,beam$^{-1}$, with the vectors of the fractional linear polarization superimposed (10\arcsec\ corresponds 
to 28\% of the fractional linear polarization). 
The size of the beam is indicated by the circles in the left upper corners of the images.}
\label{fig:pol}
\end{figure}

Maps of the linearly polarized intensity and fractional polarization
were made by combining the NVSS Stokes $Q$ and $U$ maps with the AIPS
task COMB. This allows for the determination of the polarized flux density, the 
fractional polarization, and the $E$-vectors polarization angle. The
total intensity NVSS map with the electric field $E$-vectors superimposed
is shown in Figure\,\ref{fig:pol}\,(b), while Figure\,\ref{fig:pol}\,(c) presents the linearly polarized
intensity map with the vectors of fractional linear polarization superimposed.
From these maps it can be seen that the extended structure is mildly polarized. 
The total integrated polarized flux intensity of the extended source is $7.22\pm0.36$\,mJy, 
which gives about 5\% for the mean fractional polarization. The integrated polarized flux 
density of the compact quasar core is $4.64\pm0.23$\,mJy, and the corresponding 
degree of linear polarization is therefore about 1\%. We found different orientations of the electric
vectors in different regions of the structure, as well as changes in the levels of
polarization across the source. The polarization percentage increasing along the tails (i.e.,
away from spots W and E to the south), resembles typical polarization properties of tailed radio galaxies 
(e.g., \citealt{fer98,gui08,gui10,prat13}). The polarization $E$-vectors, however, are mostly perpendicular 
to the tails. This indicates that the magnetic field vectors, in the case of negligible rotation measure (RM) values, 
would be transverse to the resolved structures, unlike in other tailed radio galaxies. Such an `unusual' property
may on the other hand simply suggest a substantial rotation of the polarization plane inside the cluster Abell\,585.
In order to determine the real distribution of the magnetic $B$-field directions, 
it is necessary to correct the radio data for the Faraday rotation, and for that purpose follow-up 
radio polarization measurements at other wavelengths with high angular resolution would be needed.
The RM of the quasar is $12\pm18$\,rad\,m$^{-2}$ \citep{d1}. 
\citet{tay09} claimed a comparable value of $38.3\pm15.9$\,rad\,m$^{-2}$
for the polarized radio feature located north-west to the quasar; this feature coincides with the diffuse
tail extending from our spot/head E. The provided value has to be taken with caution, however, since it was
estimated using only two nearby frequencies of 1365\,MHz and 1435\,MHz.

\section{Origin of the Extended Radio Structures}

\subsection{Quasar Jet?}

B3\,0727+409 is a luminous radio source with the 1400\,MHz power of $P_{\rm 1400\,MHz}=6.3 \times 10^{28}$\,W\,Hz$^{-1}$. 
Its SDSS spectrum reduced through the standard procedure (see \citealt{k1}) allows us to estimate the mass
of the central black hole $M_{\rm BH} = (3.33 \pm 1.70) \times 10^{8} M_{\odot}$ when using the MgII line, 
and $(2.26 \pm 0.34)\times 10^{8} M_{\odot}$ when using the CIV line. The resulting value
is characteristic for radio-loud quasars at high redshifts (e.g., \citealt{w2,k1}). 
Also the bolometric luminosity of the accreting matter in B3\,0727+409 estimated from the MgII line is typical for
quasars in general, namely $L_d \simeq 1.5 \times10^{45}$\,erg\,s$^{-1}$, implying high accretion rate 
$\dot{m}_{\rm acc} \simeq \eta_d^{-1} \, (L_d/L_{\rm Edd} ) \sim 0.4$ (in the Eddington units) 
for the standard $\eta_d \simeq 10\%$ radiative efficiency of the accretion disk.

As mentioned previously, B3\,0727+409 is a VLA phase calibrator, but it was also a VLBA phase calibrator.
The existing VLBA maps show one-sided nuclear jet oriented toward the extended structure under investigation.
Interestingly, the position angles of the nuclear jet and of the extended structure are similar:
the former one is 127\degr, 125\degr, 123\degr\ at, respectively, 2268.3, 4995.5 and 8338.3\,MHz, 
while the position angle of the extended structure as measured in the FIRST map between the quasar core and 
the peak of the radio emission in the western `head' [$\alpha$\,$=\rm 07^{h}30^{m}41\fs0$, $\delta$\,$=+40\degr51\arcmin43\farcs7$] 
is 134\degr. This agreement may suggest that the extended radio structure observed around the central region of Abell\,585 
is not related to the cluster, but instead represents the large-scale jet/lobe of the background quasar B3\,0727+409, as in fact
proposed in \citet{an1}. Additionally in such a case, the X-ray emission seen in the ROSAT
data likely extend along the radio structure, and could possibly be accounted for by the jet emission as well since 
large-scale quasar jets are established luminous X-ray emitters (see \citealt{hk04} for a review).

The angular distance between the quasar core and the western radio `head' is 2\farcm72, which, if the radio 
structure is at the quasar redshift indeed, corresponds to the linear size of 1.3\,Mpc. 
The de-projected physical size of the large-scale radio structure of 
B3\,0727+409 would be in such a case $> 5$\,Mpc, for the anticipated quasar viewing angle $<30$\degr and
the assumed source intrinsic symmetry. In general, lobes with linear sizes exceeding Mpc, called `giant radio sources', 
are known to be associated with radio galaxies and quasars even at high redshifts (see \citealt{k1}); the 
largest object of this kind known to date is characterized by the projected size of about 4.7 Mpc (J1420$-$0545; \citealt{m1,m2}). 
On the other hand, morphological, spectral, and polarization properties of giant radio sources are very different from those
of the extended radio structure revealed by the high-resolution radio maps discussed in this paper.
Moreover, in the framework of the quasar scenario, the source energetics would be rather extreme, as discussed below. 

The 1400\,MHz flux of the extended structure $\simeq 130$\,mJy (see \S\,4.1), if produced at the redshift of $z=2.5$, gives the 
monochromatic radio luminosity $L_{1400}\simeq 10^{44}$\,erg\,s$^{-1}$. Since only a moderate beaming for quasar radio jets is expected 
on large scales, with the corresponding jet bulk Lorentz factors  $\Gamma_j \sim 3$ \citep{wa97}, the total emitted 
radio power of the source would therefore read as $L_{\rm em}\sim \Gamma_j^{-2} f_{\rm bol} \, L_{1400} \sim 10^{44}$\,erg\,s$^{-1}$
for the bolometric correction factor $ f_{\rm bol} \sim 10$ (following from the assumption that the jet radio continuum extends from 
$\gtrsim 10$\,GHz down to MHz frequencies with the average spectral index of $\alpha \sim -0.7$). This emitted radio power would 
then require the total jet kinetic power $L_j \sim \eta_{\rm j}^{-1} \, L_{\rm em}$ in excess of $ \sim 10^{46}$\,erg\,s$^{-1}$ 
(assuming the standard $\eta_j \lesssim 1\%$ jet radiative efficiency), which is close to the maximum rate of the SMBH 
energy extraction in the system, $L_{j,\,{\rm max}} \sim 3 \, \dot{M}_{\rm acc} \, c^2 \sim 3 \times 10^{46}$\,erg\,s$^{-1}$ \citep{mck12}.
This power would have to be continuously supplied by the matter accretion for at least 
$t_{\rm life} > 100$\,Myr, in order to power the $>5$\,Mpc-scale flat-spectrum radio cavity, and this is assuming a rather high jet 
advance velocity of $\sim 0.1c$ (see the related discussion in \citealt{m2}). 

The aforementioned X-ray emission is in this context even more problematic, since the observed X-ray flux 
$\sim 10^{-12}$\,erg\,cm$^{-2}$\,s$^{-1}$, if due to the quasar jet, implies a very high luminosity 
$L_{\rm X}\simeq 5 \times 10^{46}$\,erg\,s$^{-1}$. Note that any significant beaming which could be invoked to reduce the 
total emitted power of a jet, would require at the same time very small jet viewing angle, and hence the de-projected linear
size of the structure exceeding 10\,Mpc. For all these reasons, we consider the quasar scenario for the 
analyzed extended radio structure implausible.

\subsection{Cluster Halo or Relic?}

At the redshift of the galaxy cluster Abell\,585, the observed 1400\,MHz flux of the extended radio structure
corresponds to the radio luminosity of $L_{1400}\simeq 7 \times 10^{40}$\,erg\,s$^{-1}$, which is in the range
of radio luminosities for typical cluster radio halos \citep{ct02,ferrari08,feretti1}. 
However, cometary-like morphologies of two distinct features with clearly defined `hotspots' or `heads' 
revealed on the high-resolution radio maps, their relatively small sizes $\lesssim 200$\,kpc (if at $z = 0.12$), 
$\sim 5\%$ radio polarization, and relatively flat radio spectra ($\alpha \sim -0.7$), are all in a disagreement with such an association.

As we mentioned in \S\,2, Abell\,585 is a member of the SCL\,074 supercluster which may represent a dynamic merging structure. The energy dissipated in large-scale shocks during cluster-cluster interactions can be channeled into relativistic electrons and magnetic field (e.g. \citealt{vW12}), resulting in a formation of extended radio structures similar to those seen in e.g. Abell\,85 \citep[see \S\,6 below]{Slee01}. Yet the analyzed radio features can be hardly identified with any sort of radio relics, since those elongated or even filamentary radio structures with sizes $0.1 - 1$\,Mpc are found mostly at the outskirts of galaxy clusters, and, even though polarized, are characterized by very steep radio spectra ($\alpha \leq -1$; see, e.g., \citealt{vW10}).

\subsection{Is the structure caused by individual galaxy from the cluster?}

Radio-loud AGN typically found in clusters of galaxies are low-power radio galaxies, whose jets and lobes 
are often distorted and bent, with the characteristic U- or C-shape large-scale morphologies. 
Such sources are denoted as narrow-angle tail (NAT) and wide-angle tail (WAT) radio galaxies, respectively. 
The dramatic bending of the jets is due to a combination of a parent galaxy motion within the cluster, and local 
pressure gradients in the ICM. For the early comprehensive and statistical study of NAT sources see, e.g., 
\citet{o1} and \citet{o2}. The two radio features resolved in our high-resolution radio maps,
and especially the western one, do resemble NAT sources taking into account their morphologies, location around
the central parts of a dynamical cluster, radio powers ($L_{1400}\sim 3 \times 10^{40}$\,erg\,s$^{-1}$ each), and 
finally linear sizes of the order of hundreds of kpc. The main problem with this interpretation is the lack of any obvious 
galaxies which could be identified as hosts of the two radio structures: as already mentioned above,
no optical counterparts can be found in the SDSS data at the positions of the two radio `heads', which presumably would
mark the positions of any putative radio cores. Moreover, none of the galaxies located within the central parts of the cluster Abell\,585 
in close proximity to the radio features reveal any signatures of the AGN activity (see Table\,1).
We have verified this finding by a careful examination of the all-sky mid-infrared (MIR) data provided by the 
NASA Wide-field Infrared Survey Explorer satellite (WISE; \citealt{w3})\footnote{The point-spread function of the WISE telescope 
in four different filters corresponds to a $\leq 12$\,\arcsec\ Gaussian; nominal 5$\sigma$ point source sensitivity from 
$\sim$0.08 up to 6\,mJy; see also \texttt{http://wise2.ipac.caltech.edu/docs/release/allsky}} in the range $3.4-22$\,$\mu$m.
Indeed, all the MIR emitters found in the region of interest correspond to the optically-detected galaxies and the 
quasar B3\,0727+409, and therefore we can exclude a possibility for the presence of heavily-obscured AGN
at the positions of the radio cores of the two analyzed radio features.

But can the two resolved radio features be associated with background radio galaxies located at redshifts high enough
that their hosts are simply below the flux limit of the SDSS? Using the $r$-band Hubble diagram for 3C galaxies given in \citet{snell96}, 
and the detection limit in the $r$-band SDSS images of $\rm 22.5^{m}$ \citep{zeh11}, we estimate the minimum redshift of 
putative parent galaxies in this scenario as $z \gtrsim 1.7$. At such high redshifts, the angular size of the radio tails ($\sim2\farcm5$) would correspond to the linear size of $\gtrsim 1.3$\,Mpc. Keeping in mind that the largest head-tail radio galaxy known to date, 
peculiar 3C\,129 at $z=0.021$, is only twice as long \citep{jeg83}, we consider the possibility of two neighboring distant giant 
tailed radio sources to be very low.

The two cluster member galaxies closest to the cores of the analyzed radio structures with projected separations of about 
$\gtrsim 50$\,kpc are the giant but otherwise standard elliptical with no prominent emission lines, and the 
bright face-on spiral with bright optical core (lacking however any spectroscopic redshift determination), designated 
respectively as `P' and `S' in Figure\,\ref{fig:perley} (see also Table\,1). Let us therefore consider a scenario in which 
the radio structures in question are due to the jet activity of SMBHs which experienced gravitational wave recoil kicks 
in galaxy merging process within the central parts of Abell\,585. The gravitational recoil of merging black holes is widely considered in the 
literature, typically in the context of the cosmological evolution of SMBHs and the galaxy formation (e.g., \citealt{mad04,merritt,bak06}). Candidate AGN related to this process have been identified based 
on peculiarities in their line emission \citep{kom08,sh09,rob10}, or the presence of 
double nuclei separated by projected distances of a few up to several kpc \citep{com09C,jon10,kee11}.

If the gravitational recoil process is of any relevance for the systems analyzed in this paper,
the recoil kicks have to be strong enough for a SMBH to be able to escape the gravitational potential of a
massive host galaxy (see in this context \citealt{km08}). Recent numerical simulations indicate 
that such large recoil velocities exceeding 1000\,km\,s$^{-1}$
are possible under specific conditions of rapidly spinning black holes with comparable masses and a particular orientation
of their spin vectors at small angles to the orbital plane (e.g., \citealt{Cam07,gon07,bak08}).
And in fact dry mergers of massive ellipticals in central parts of galaxy clusters may provide the required conditions, since
such systems are poor in cold gaseous content (while this is the abundant cold gas which is believed to enable an 
efficient alignment of black hole spins perpendicular to the orbital plane during the merger; see \citealt{bog07}), 
but on the other hand to contain rapidly spinning SMBHs on average (see in this context, e.g., \citealt{vol07,vol13}). One could speculate further that the recoiled SMBHs traveling through the ICM 
after leaving host galaxies accrete only the hot intracluster gas at low (Bondi) rates, and as such form radiatively inefficient 
accretion disks; this could potentially explain the apparent lack of any AGN activity at MIR-optical wavelengths at the 
position of the cores of the analyzed radio structures.

Yet, on the other hand, the probability of having two major mergers in the central parts of a single cluster, within the period 
overlapping with radiative lifetimes of the two radio galaxies ($<100$\,Myr; see \S\,4.2), is very low. There are also no obvious 
`remnant' pairs of elliptical hosts which should be expected around the positions of the E and W spots in the framework of
the gravitational recoil scenario: as emphasized above, only one giant elliptical and one bright spiral are detected
within the radius of $\sim 100$\,kpc from the radio cores. Note in this context that powering the lobes with the observed
radio luminosities of $L_{1400}\sim 3 \times 10^{40}$\,erg\,s$^{-1}$ requires jet kinetic luminosities of the order of, at least, 
$L_j \sim 10^{43}$\,erg\,s$^{-1}$, and therefore very high black hole masses for the anticipated low/moderate accretion rates.
Hence any association of the discussed radio structures with some low-mass (and as such undetected) hosts seems 
implausible. In order to justify this statement, we note that with the Bondi accretion rate $\dot{M}_{\rm B} = 4 \pi \, n_g \, 
m_p \, G^2 M_{\rm BH}^2 \, c_s^{-3}$ and the following maximum jet kinetic luminosity $L_j \sim \dot{M}_{\rm B} c^2$,  
one obtains the limiting black hole mass
\begin{eqnarray}
& M_{\rm BH} & \simeq 10^9 \, M_{\odot} \times \\ 
&& \left({L_j \over 10^{43}\,{\rm erg/s}}\right)^{1/2} \left({n_g \over 0.01\,{\rm cm^{-3}}}\right)^{-1/2}
\left({T_g \over 5 \times 10^{7}\,{\rm K}}\right)^{3/4} \nonumber
\end{eqnarray}
where $n_g \sim 0.01$\,cm$^{-3}$, $T_g \sim 5 \times 10^7$\,K, and $c_s = (5 k T_g / 3 m_p)^{1/2}$ are respectively 
the gas number density, temperature, and sound velocity expected for the central parts of a galaxy cluster.

\section{Summary} 

The peculiar arcmin-scale radio structure observed in the direction of the central parts of the galaxy cluster Abell\,585 ($z=0.12$)
is unassociated with the distant ($z=2.5$) radio quasar B3\,0727+409, as we demonstrated here by means of a careful
analysis of all the available radio data for the system. This structure consists of two dominant features which resemble typical 
lobes of cluster radio galaxies of the NAT type. However, we are not able to identify host galaxies of the two features in the 
available optical (SDSS) and infrared (WISE) surveys. We speculate if the analyzed systems are examples of extreme
gravitational recoils of SMBHs in the process of merging massive ellipticals within the central parts of a rich cluster environment,
but we do not reach any robust conclusions regarding the origin of the detected radio features.

Large-scale radio structures lacking obvious host galaxies may not be that unique for the local clusters. For example,
radio maps of B\,1753$+$580 found at the position of Abell\,2289 \citep{o0} show two edge-dimmed 
tails with bright `heads' and with a $z=0.224$ galaxy located between them \citep{ow1}, but without any 
detectable compact radio core. \citet{Slee98} presented a detailed study of the radio structure in cluster Abell\,4038 
resembling the evolved lobes of a luminous radio galaxy, but with the most likely host elliptical galaxy displaced by about 
20\,kpc with respect to the central parts of the radio structure. \citet{Slee01} discussed three other possibly analogous 
systems in clusters Abell\,13, Abell\,85, and Abell\,133. Inspecting FIRST maps of several galaxy clusters from 
\citet{koe07}, we add the case of Max BCG J130.34220+61.21246 ($z=0.13$) to the literature examples,
and note the striking similarity between the radio structures discussed in this paper and those in the $z = 0.16$ 
cluster MaxBCG J250.34552+38.03597 that lack any optical identifications.
There may be more examples of such peculiar structures awaiting discovery with high-resolution 
multiwavelength radio observations of lesser-known clusters of galaxies in formation. 

\section{Acknowledgments} 
We thank the anonymous reviewer for comments which have significantly improved the paper.
This research has made use of the NASA/IPAC extragalactic database (NED), which is operated
by the Jet Propulsion Laboratory, Caltech, under contract with the National Aeronautics
and Space Administration. We acknowledge use of the Sloan Digitized Sky Survey. 
\L .S., M.O. and M.J. are supported by Polish NSC grants DEC-2012/04/A/ST9/00083 
and DEC-2013/09/B/ST9/00599, respectively. Work by C.C.C. at NRL is supported 
in part by NASA DPR S-15633-Y.

\label{lastpage}

\end{document}